\documentstyle[prb,aps,epsf]{revtex}
\begin{document}
\draft

\twocolumn[\hsize\textwidth\columnwidth\hsize\csname@twocolumnfalse\endcsname

\title{Mode-locking in driven vortex lattices with transverse ac-drive
and random pinning}

\author{Alejandro~B.~Kolton$^1$,
Daniel~Dom\'{\i}nguez$^1$, and Niels~Gr{\o}nbech-Jensen$^{2,3}$}
\address{
$^1$Centro Atomico Bariloche and Instituto Balseiro, 8400 S.~C.~de~Bariloche, Rio Negro, Argentina\\
$^2$Department of Applied Science, University of California, Davis, California 
95616\\
$^3$NERSC, Lawrence Berkeley National Laboratory, Berkeley, California 94720}

\date{\today}
\maketitle
\begin{abstract}
We find mode-locking steps in simulated current-voltage characteristics of
driven vortex lattices with {\it random} pinning when an applied 
ac-current is {\it perpendicular} to the dc-current. For low 
frequencies there is mode-locking only above a non-zero threshold
ac force amplitude, 
while for large frequencies there is mode-locking for any small ac force.
This is consistent with the nature of {\it transverse} temporal order 
in the different regimes in the absence of an applied ac-drive.
For large frequencies the magnitude of the fundamental mode-locked step
depends linearly with the ac force amplitude.
\end{abstract}

\pacs{PACS numbers: 74.60.Ge, 74.40.+k, 05.70.Fh}

]                

\narrowtext

Non-linear dynamics of vortices driven by a current in random media
leads to several interesting non-equilibrium phases, such as plastic flow,
moving smectic and moving Bragg glass.
\cite{plastic,theory,EXP,SIM,kolton,trc,BF}
These dynamical phases can
be characterized by their temporal
order\cite{theory,SIM,BF,troya,togawa} and mode-locking
responses.\cite{fiory,harris,mlock}
When a vortex array with average intervortex spacing
$a$ is moving at a high enough velocity $v$ ,
it is possible  to  have temporal
order at the washboard frequency $\omega_0=2\pi v /a$, which results in
a peak at $\omega_0$ in the voltage power
spectrum.\cite{troya,togawa} This has
been observed in numerical simulations\cite{SIM,mlock} and 
also recently in experiments.\cite{troya,togawa}
When the system is driven by a dc + ac force with frequency $\Omega$,
interference phenomena leads to mode-locking steps for vortex velocities
such that $\omega_0=(p/q)\Omega$.\cite{fiory,harris,mlock} 
This interesting effect has been observed experimentally
by Fiory\cite{fiory} and by Harris {\it et al}.\cite{harris} 
Recently, we have numerically studied
how the existence of mode-locking in driven vortex lattices
depends on the presence of temporal order in each dynamical
regime.\cite{mlock}  

Mode-locking phenomena has been extensively studied in other systems
in the past, e.g., Josephson junctions (Shapiro steps),\cite{shapiro} 
Josephson junction arrays,\cite{jja} 
superconductors with periodic pinning\cite{periexp,charles,square} 
and charge density waves (CDW).\cite{cdw,cdw2}
Driven vortex lattices with random pinning have two important
features that distinguish them from these systems.
(i) There is no inherent periodicity, as for example in Josephson
junction arrays and superconductors with periodic pinning. 
Temporal order and periodicity are induced dynamically due to the
vortex-vortex interaction, which tends to favor a structure close
to a triangular vortex lattice at large velocities.\cite{theory} 
(ii) The vortex displacements are two-dimensional vectors. This is an
important difference with respect to CDW systems where the displacement
field is a scalar.\cite{cdw2} In particular, the behavior 
of the displacements in the direction perpendicular
to the driving force shows phenomena like a transverse critical 
current\cite{theory,SIM,trc} and a transverse
freezing transition\cite{theory,kolton} at high velocities.
It can therefore be interesting to study the possibility of mode-locking when 
an ac force is applied in the direction {\it perpendicular} to the
direction of the dc driving force. Recently, 
it has been found in rectangular periodic pinning arrays\cite{square}
and in Josephson junction arrays\cite{tjja} that a transverse ac force
leads to a new type of ``transverse'' phase-locking in these cases.
In this paper we will investigate the possibility of {\it transverse
mode-locking} in driven vortices with random pinning.

The dynamics of a vortex in position ${\bf r}_i$ is given
by:\cite{SIM,kolton}
\begin{equation}
\eta \frac{d{\bf r}_i}{dt} = -\sum_{j\not= i}{\bf\nabla}_i U_v(r_{ij})
-\sum_p{\bf \nabla}_i U_p(r_{ip}) + {\bf F}(t),
\end{equation} 
where $r_{ij}=|{\bf r}_i-{\bf r}_j|$ is the distance between vortices $i,j$,
$r_{ip}=|{\bf r}_i-{\bf r}_p|$ is the distance between the vortex $i$ and
a pinning site at ${\bf r}_p$, $\eta=\frac{\Phi_0H_{c2}d}{c^2\rho_n}$ is the
Bardeen-Stephen friction and ${\bf F}(t)=
\frac{d\Phi_0}{c}[{\bf J}_{dc}+{\bf J}_{ac}\cos(\Omega t)]\times{\bf z}$
is the driving force due to an alternating current 
${\bf J}_{ac}\cos(\Omega t)$
superimposed to a constant  current ${\bf J}_{dc}$.
The vortex-vortex interaction is considered logarithmic:
$U_v(r)=-A_v\ln(r/\Lambda)$, with $A_v=\Phi_0^2/8\pi\Lambda$,
and $\Lambda=2\lambda^2/d$ is the effective penetration depth
of a thin film of thickness $d$.\cite{kolton,mlock}
The vortices interact with a random distribution of
attractive pinning centers with 
$U_p(r)=-A_p e^{-(r/\xi)^2}$, $\xi$ being the coherence length. 
Length is normalized by $\xi$, energy by $A_v$, 
and time by 
$\tau=\eta\xi^2/A_v$.  We consider $N_v$ vortices and $N_p$ pinning
centers in a rectangular box of size $L_x\times L_y$, 
and the normalized  vortex density is $n_v=N_v\xi^2/L_xL_y=B\xi^2/\Phi_0$.
Moving vortices induce a total electric field  ${\bf
E}=\frac{B}{c}{\bf v}\times{\bf z}$, with ${\bf v}=\frac{1}{N_v}\sum_i 
{\bf v}_i$. 

We study the response of the vortex lattice to a dc 
force plus a {\it transverse} ac force,
${\bf F}=F_{dc}{\bf y}+ F_{ac}\cos(\Omega t){\bf x}$ 
 solving Eq.~(1) for different values of $F_{ac}$ and $\Omega$.
The simulations are at $T=0$ for a vortex density $n_v=0.04$ in
a box with $L_x/L_y=\sqrt{3}/2$, and 
$N_v=64,100,144,196,256,400$ (we show results for $N_v=256$), and
we consider weak pinning strength of $A_p/A_v=0.05$
with a density of pinning centers being $n_p=0.08$.
We impose periodic boundary
conditions and the long-range
interaction is determined by Ref.~\onlinecite{log}.
The time integration step is $\Delta t=0.001\tau$ and averages are
evaluated during $131072$ steps after $3000$ steps for 
equilibration. 

In a previous work\cite{mlock} we studied the case of 
a longitudinal ac force,  relating the mode-locking response
with the presence of temporal order for the longitudinal component
of the velocity. 
Here we analyze now the {\it transverse temporal order} from  
the transverse power voltage spectra (corresponding to the transverse
velocity), which are
\begin{figure}
\centerline{\epsfxsize=8.5cm \epsfbox{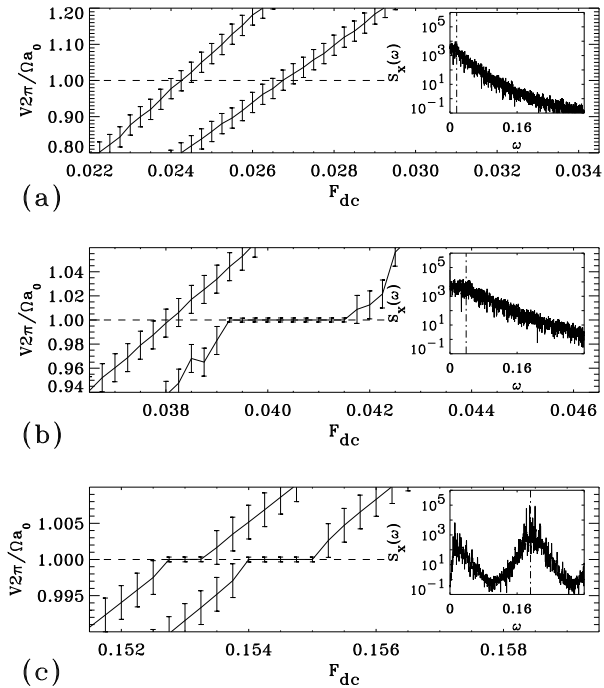}}
\caption{
Velocity-force curve  around the main
interference condition $V = \Omega a_0/2 \pi$ for three typical 
drive frequencies $\Omega$. Each case show results 
for two values of amplitude $F_{ac}$ 
(the curves are shifted in $F_{dc}$ for clarity). 
Insets show corresponding voltage power spectrum for 
$F_{ac}=0$ and $V \approx V_{step}$. Vertical dashed line in the 
spectral density indicates the expected washboard frequency $\omega_0$.
(a) $\Omega=0.02$, $F_{ac}=0.01$ (left), $F_{ac}=0.03$ 
(right). 
(b) $\Omega=0.04$, $F_{ac}=0.02$ (left), 
$F_{ac}=0.08$ (right).
 (c) $\Omega =0.19$, $F_{ac}=0.09$ (left), 
$F_{ac}=0.23$ (right).}
\end{figure}\noindent
shown in the insets of Fig.~1. 
They are calculated  as 
$S_x(\omega)=|\frac{1}{T}\int_0^Tdt V_x(t)\exp(i \omega t)|^2$
at the different dynamical regimes for $F_{ac}=0$.\cite{kolton}
The first regime above  the critical
depinning force $F_c$ is the plastic flow regime
($F_c < F_{dc}< F_p$, $F_c \approx 0.01$,  $F_p \approx 0.03$).
In this case we find a broad band spectrum without temporal
order [inset of Fig.~1(a)]. Similar behavior is found in the
``smectic flow'' regime ($F_p < F_{dc}< F_t$, 
$F_t\approx0.06$), shown in the inset of Fig.~1(b).
This is reasonable, since we know that the transverse
motion is diffusive in both regimes.\cite{kolton} 
Only for $F_{dc} > F_t$, in the ``transverse solid''
regime, we find clear evidence of temporal order 
in the transverse velocity. This is seen in the inset of Fig.~1(c) where
well developed peaks appear at the washboard frequency,
$\omega_0$, and its harmonics. 
We are now ready to study the response
to a superimposed transverse ac-force $F_{ac}\cos(\Omega t)$, for 
varying values of $F_{ac}$. For a given $\Omega$,
we expect the main interference step ($p=q=1$) to occur when
$V=V_{step}=\Omega a/2 \pi$  ({\it i.e.}, $\Omega=\omega_0$) if there
is mode-locking. 
We therefore choose the values of $\Omega$ 
such that the expected step, $V_{step} = \Omega a/2\pi$,
would correspond to velocities $V$ belonging 
to a given dynamical regime 
of the limit $F_{ac}=0$. Each simulation is started at
$\langle v_y \rangle \approx 0.975 \Omega a/2 \pi $ 
with an ordered triangular 
\begin{figure}
\centerline{\epsfxsize=8.5cm \epsfbox{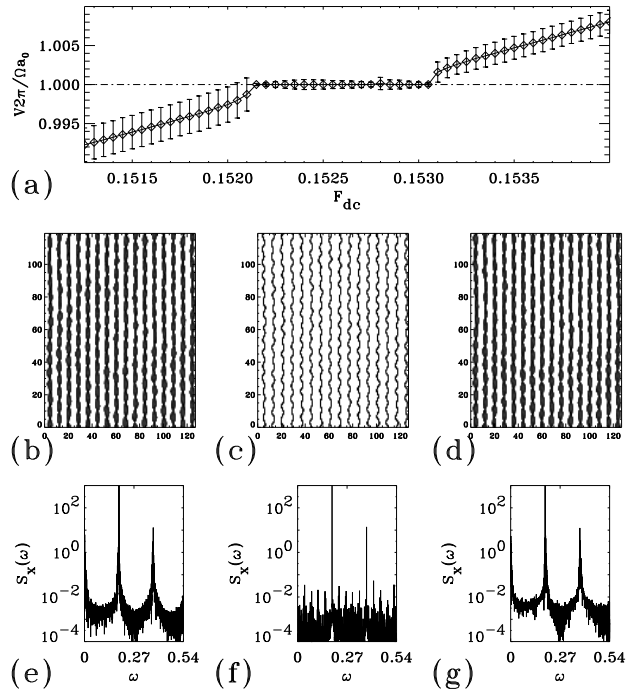}}
\caption{(a) Velocity-force curve  around the main
interference step for $\Omega=0.08$ and $F_{ac}=0.3$. 
(b)-(c)-(d) Typical time averaged coarse-grained 
density of vortices for a mode-unlocked state bellow the step, 
a mode-locked state in the step, and a mode-unlocked state above 
the step, respectively.
(c) Typical voltage power spectrum for the three ac-driven 
regimes mentioned above.}
\end{figure}\noindent
lattice up to values such that 
$\langle v_y \rangle \approx 1.025 \Omega a/2 \pi$ 
by slowly increasing the dc force $F_{dc}$ with 
$\Delta F_{dc}= 0.00005-0.00025$. 
For low $\Omega$, for which
we have plastic flow when  $F_{ac}\rightarrow0$, we
find that there are no interference steps in a wide range of  $F_{ac}$
(shown in Fig.~1(a) for $F_{ac}/V_{step}<1$ (left curve) and 
$F_{ac}/V_{step}>1$ (right curve)).
For intermediate $\Omega$, for which  we
have smectic  flow when $F_{ac}\rightarrow0$, 
we find that there are no steps
for small amplitudes, $F_{ac}/V_{step}<1$, while there are steps for
$F_{ac}/V_{step}>1$, as shown in Fig.~1(b) in  the left and right curves,
respectively. 
For high $\Omega$, corresponding  to a  transverse solid regime 
when $F_{ac}\rightarrow0$  we find that  there are steps both for small
$F_{ac}/V_{step}<1$ and large $F_{ac}/V_{step}>1$ values of the  ac
amplitude, as we can observe in Fig.~1(c).
We therefore find a behavior similar  to the case of longitudinal
ac forces.\cite{mlock} In the present case,
when the dynamical regime has transverse temporal order, any small 
amount of $F_{ac}$ will induce transverse mode-locking, while for
the dynamical regimes that do not have transverse temporal order,
a non-zero (threshold) value of $F_{ac}$ is needed to induce transverse
mode-locking.

In Fig.~2(a) we show in detail a typical $V-F_{dc}$ curve around the 
transverse mode-locking step. 
To visualize the spatial structure of trajectories in the transition we
define a coarse-grained vortex density  $\rho_v({\bf r},t)$. We
take a coarse-graining  scale $\Delta r < a_0$. In Fig.2(b)-(c)-(d) we 
\begin{figure}
\centerline{\epsfxsize=8.5cm \epsfbox{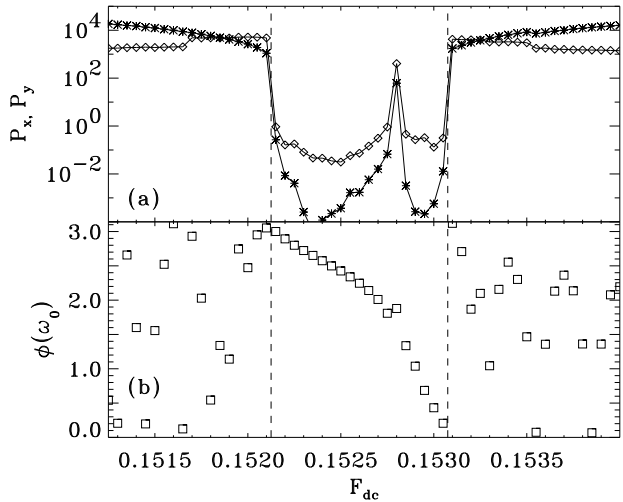}}
\caption{(a) Low frequency voltage noise in the transverse and longitudinal 
direction around the main interference step. 
The dashed lines indicate the mode-locking transitions.
(b) Phase of the washboard frequency component of the 
longitudinal voltage Fourier transform around the step.}\end{figure}\noindent
show the temporal average  $\langle \rho_v({\bf r},t) \rangle$ of the
density for three typical values of $F_{dc}$, corresponding to voltages 
$V<V_{step}$ (Fig. 2(b)), $V=V_{step}$ (Fig.2(c)), 
and $V>V_{step}$ (Fig.2(d)). We observe in Fig.2(b) that within the
mode-locked state vortices follow one-dimensional trajectories. The
wavy nature of the trajectories is, of course, due to the transverse
ac force. Figures 2(e)-(f)-(g) show typical transverse voltage spectral 
densities $S_x(\omega)$ for the three cases mentioned above.  We see
that there is a significant reduction in the width of the washboard peak
within the step; a typical signature of mode-locking.\cite{cdw,cdw2}
In Fig.3(a) we show the low 
frequency voltage noise in both directions, perpendicular $P_x$ and 
longitudinal $P_y$ to the dc-force, defined as 
$P_{x,y}=\lim_{\omega \to 0}S_{x,y}(\omega)$. 
We see that also the low frequency noise is 
greatly reduced within the step.  (There is a noise peak inside
the step which corresponds to a transition between  different
mode-locked structures).
In Fig. 3(b) we show the phase 
$\phi(\omega_0)$ of the washboard frequency component of 
the longitudinal voltage 
Fourier transform $\tilde V(\omega_0)$, defined 
as $\tilde V(\omega_0)=\sqrt{S_y(\omega_0)}\exp(i \phi(\omega_0))$.
Here we see explicitly that 
within the ``phase-locked'' state there is a well defined ``phase''
which varies within the range $0\le\phi\le\pi$. 

In Fig.~4(a-b) we show the range (width)
$\Delta F_{dc}$ for the case  $F_p < F_{dc} < F_t$ and $F_t < F_{dc}$
respectively. The error bars and the mean values  were estimated by
repeating 
\begin{figure}
\centerline{\epsfxsize=8.5cm \epsfbox{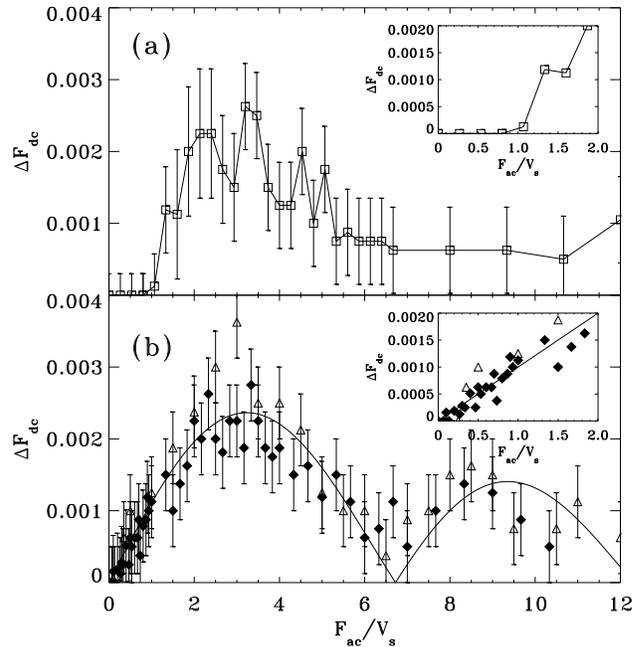}}
\caption{ Step width $\Delta F_{dc}$ vs $F_{ac} 2\pi/\Omega a_0=F_{ac}/
V_{step}$. (a) $\Omega = 0.04$. (b) $\Omega= 0.13$ 
($\triangle$) points and $\Omega= 0.19$ ($\Diamond$) points. Solid line
shows a fit to  $A|J_1(F_{ac}/V_{step}\sqrt{3})|$.}  
\end{figure}\noindent
the simulation for three different disorder realizations. In
Fig.~4(a) we show $\Delta F_{dc}$ for $\Omega=0.04$ vs $F_{ac}$, which
corresponds to the  smectic flow regime for $F_{ac}\rightarrow0$. 
We see that there is mode-locking only above a non-zero threshold value
$F_{ac}/V_{step} \approx 1$ (see inset of Fig.~1(a)). 
In Fig.~4(b) we show  $\Delta F_{dc}$ for two frequencies 
$\Omega=0.13, 0.19$ vs $F_{ac}$,  
which correspond  to the transverse solid in the
$F_{ac}=0$ limit. 
We can collapse (approximately) both
curves into a single curve if we plot $\Delta F_{dc}$  vs
$F_{ac}/V_{step}$. Our results follow closely a dependence
of the form $\Delta F_{dc} \approx A
|J_1(F_{ac}/V_{step}\sqrt{3})|$ with $A$ being a
constant. In the inset 
of Fig.~4(b) we can see that there is a linear dependence of the 
mode-locking intensity with $F_{ac}$. 
This is very different from transverse mode-locking in periodic
pinning systems,\cite{square,tjja} in which the step width follows
$\Delta F_{dc} \propto (F_{ac})^2$. 
The rather surprising result that in the random pinning case
the transverse mode-locking intensity has a linear $F_{ac}$-dependence
can be explained as a consequence of the 
existence of transverse temporal order in the $F_{ac}=0$ limit. 
We can show this with a very simple effective model. 
The moving lattice can be described approximately by an 
equation of motion for the velocity ${\bf v}$ of 
its center of mass,
\begin{eqnarray}
{\bf v}={\bf F_{dc}}+{\bf F_{ac}}\cos(\Omega t) 
-\sum_{\bf G}{\bf G}U_{\bf G}\sin({\bf G}\cdot{\bf r})\;.
\end{eqnarray}
The $U_{\bf G}$ are the components of an effective 
periodic force, due to the interaction of the nearly 
periodic moving lattice (with reciprocal vectors ${\bf G}$) 
with disorder. For weak disorder (small $U_{\bf G}$) a 
first order correction can be obtained assuming that
in zero order ${\bf r}={\bf r_0}+\langle {\bf v} \rangle t+{\bf F_{ac}}
\sin(\Omega t)/\Omega$.
This gives for the instantaneous velocity ${\bf v}$ and average 
velocity $\langle {\bf v} \rangle$ the 
following expressions at first order in $F_{ac}$:  
\begin{eqnarray}
{\bf v} &=& {\bf F_{dc}}-\sum_{\bf G}{\bf G}U_{\bf G} 
\sin\biggl[{\bf G}.\biggl({\bf r_0}+\langle {\bf v} \rangle t \biggr)\biggr] 
\nonumber \\ 
&& \biggl[J_0\bigg(\frac{{\bf G}.{\bf F_{ac}}}{\Omega}\bigg) 
+ 2 J_1\bigg(\frac{{\bf G}.{\bf F_{ac}}}{\Omega}\bigg)
\sin(\Omega t) \biggr] 
\end{eqnarray} 
\begin{eqnarray}
\langle {\bf v} \rangle &=& {\bf F_{dc}}-\sum_{\bf G}{\bf G}U_{\bf G}
\sin({\bf G.r_0})\Biggl\{  
J_0\bigg(\frac{{\bf G}.{\bf F_{ac}}}{\Omega}\bigg) 
\delta({\bf G.\langle v \rangle}) \nonumber \\ 
&-& 2J_1\bigg(\frac{{\bf G}.{\bf F_{ac}}}{\Omega}\bigg)
\delta({\bf G.\langle v \rangle} - \Omega)
\Biggr\} 
\end{eqnarray} 
We consider now an anisotropic 
triangular lattice with one of its principal axes parallel
to ${\bf F_{dc}}={\bf y}F_{dc}$, and for 
simplicity we keep only the shortest reciprocal vectors 
$\{{\bf G}\}=\{ {\bf g_s}, {\bf g_l}, {\bf g_l-g_s}  \}$
where
${\bf g_s}= {\bf x} \frac{2\pi}{a_0}\frac{2}{\sqrt{3}}$ and
${\bf g_l}= {\bf x} \frac{2\pi}{a_0}\frac{1}{\sqrt{3}}+
{\bf y} \frac{2\pi}{a_0}$. Then, we consider $
U_{\bf g_l}=U_l=U_{\bf g_l-g_s}+\Delta U_l$ and $
U_{\bf g_s}=U_s$.
Here $\Delta U_l$ represents a small deformation of the perfect triangular 
lattice in the {\bf y} direction. $U_s$ and $U_l$ could be related,
respectively, to the smectic and longitudinal
structure factor peaks of the moving vortex system. 
For ${\bf F_{ac}} = F_{ac} {\bf x}$ we obtain
for the transverse velocity in the limit $F_{ac}=0$,
\begin{equation} 
v_x = 
-\frac{2\pi \Delta U_l}{a_0\sqrt{3}} \sin \Bigl[ \frac{2\pi}{a_0} 
(r_0+\langle v\rangle t) \Bigr]\;. 
\end{equation}
With this approach we obtain the phase-locking range for the 
first interference step $\Omega=\omega_0$,  
\begin{equation}
\Delta F_{dc}  = \frac{4\pi |\Delta U_l|}{a_0} 
|J_1(\frac{{\bf g_l}.{\bf F_{ac}}}{\Omega})| 
=  \frac{4\pi |\Delta U_l|}{a_0} 
|J_1(\frac{F_{ac}}{V_{ \rm step}\sqrt{3}})|\;.
\end{equation}
Even when it was derived for small $F_{ac}$ and small disorder,  
Eq.~(6) is the approximate relation found in the 
simulation, shown in Fig. 4(b). From equations 
(5) and (6) we see that, for small $F_{ac}$,
temporal order in the transverse direction is directly related
through $\Delta U_l$ 
with the linear dependence of $\Delta F_{dc}$ on $F_{ac}$.
It is interesting to note that in the perfectly periodic case,
$\Delta U_l=0$, there is no ``transverse temporal order'' and
the mode-locked step widths would be quadratic in $F_{ac}$.  This is because
in the periodic case vortices would move in straight lines without 
any transverse component of the velocity, and transverse mode-locking
would arise as a second order effect. In conclusion,
a small amount of disorder (random pinning) is enough to 
induce transverse temporal order, and, thus,
a linear step width dependence with $F_{ac}$.

We acknowledge discussions with V.~I.~Marconi and C. Reichhardt. 
This work  has been supported by 
CONICET, CNEA  and ANPCYT (Argentina) and by
the Director, Office of
Adv. Sci. Comp. Res., 
Div. of Math. Inf. and
Comp. Sci. of the U.S.D.O.E. (contract
DE-AC03-76SF00098).

\end{document}